\documentclass[12pt]{article}
\usepackage{amsfonts}
\usepackage{epsfig,amsmath,amssymb,amsfonts}

\newcommand{\be}{\begin{equation}}
\newcommand{\ee}{\end{equation}}
\newcommand{\ba}{\begin{array}}
\newcommand{\ea}{\end{array}}

\newcommand{\qe}{\end{equation}}

\begin{document}

\centerline{\large\bf
Photon Induced Entanglement in Atom-Cavity Systems}
\vspace{4ex}

\begin{center}
Li-Li Lan$^{1}$, Xiang-Bin Wang$^{2}$ and Shao-Ming Fei$^{1}$

\vspace{2ex}

\begin{minipage}{5.5in}

\small $~^{1}$ {\small School of Mathematical Sciences, Capital
Normal University, Beijing 100048}

{\small $~^{2}$ Department of Physics, Tsinghua University, Beijing 100084}

\vspace{2ex}

\centerline{\large Abstract}
 \vspace{1ex} We study the evolution of
quantum entanglement in double cavity systems. The entanglement of
cavity atoms induced by entangled pair of photons is investigated.
Both entanglement sudden death and entanglement sudden birth
phenomena are shown to be existed and analyzed in detail. We also
propose a strategy to enhance the entanglement between the atom in one cavity
and the photon in another cavity by using quantum Zeno effect.
\smallskip
\smallskip
\end{minipage}
\end{center}
\bigskip

\section{Introduction}

Entanglement plays a key role in quantum computation and quantum information
processing \cite{1}.
Due to the interactions with the environment
in preparation and transmission, the entangled states usually become
mixed ones that are no longer maximally entangled.
It is of great importance to
known and control the evolution of entanglement in quantum systems.
The entanglement evolution under the influence of
local decoherence has been studied by many authors recently
\cite{thomas,lzg,yuting,dodd,karol,frank}.
For a bipartite system with one subsystem undergoing an arbitrary
noisy channel, elegant relations have been obtained
between the concurrence of the initial and final states
\cite{thomas,lzg}. In \cite{yuting} the authors
investigated the time evolution of entanglement of a
bipartite qubit system undergoing various modes of decoherence.
It is found that although it
takes infinite time to complete the decoherence locally,
the global entanglement may vanish in finite time, a
phenomenon so called entanglement sudden death (ESD).
Such phenomena have been studied further
in various systems with different
entanglement measures and purposes \cite{pp}.
Experimental evidences of ESD have been also reported for optical
setups and atomic ensembles \cite{p30}.

In this paper we study double cavity systems with a two level atom
in each cavity. In stead the case that the atoms are initially entangled
\cite{yuting}, we consider that the atoms are initially spatially
separated and in a separable state. We investigate the evolution of
quantum entanglement when a pair of entangled photons are introduced
into the cavities. It is found that when two atoms are initially in
the ground state, there exits a kind of entanglement transfer between
atoms and photons. When two atoms were initially in exited state,
then there exit both entanglement sudden death and entanglement
sudden birth (ESB) phenomena between the two atoms, and between the atom in one cavity
and the photon in another cavity. For both initial conditions we find that the
entanglement between the atom in one cavity and the photon in another cavity is rather small in general.
We show that their entanglement can be enhanced in terms of quantum Zeno effect.

We use concurrence as the measure to characterize the quantum entanglement of a two-qubit state
$\rho$ \cite{7},
 $$
 C(\rho)=\max\{0,\sqrt{\lambda_{1}}-\sqrt{\lambda_{2}}-\sqrt{\lambda_{3}}-\sqrt{\lambda_{4}}\},
 $$
 where $\lambda_{i}$ are the eigenvalues, in decreasing order, of the matrix
 $\rho(\sigma_{y}\otimes\sigma_{y})\rho^{\ast}(\sigma_{y}\otimes\sigma_{y})$,
$\rho^{\ast}$ denotes the complex conjugation of $\rho$ and $\sigma_{y}$ is the Pauli matrix. If a
density matrix $\rho$ only contains nonzero elements along the main
diagonal and anti-diagonal such as
\be\rho=\left(
\begin{array}{cccc}
a&0&0&\omega\\
0&b&z&0\\
0&z^{\ast}&c&0\\
\omega^{\ast}&0&0&d
\end{array}
\right),
\ee
then its concurrence is verified to be of the form
\begin{equation}
C(\rho)=2\max\{0,|z|-\sqrt{ad},|\omega|-\sqrt{bc}\}.
\end{equation}

For qubit-qutrit systems, there is no analytical formula of concurrence in general.
We use negativity \cite{9,10} as the measure of quantum entanglement, which also gives rise to
a necessary and sufficient criterion for separability of qubit-qubit or qubit-qutrit
states. The negativity $N(\rho)$ of a state $\rho$ is defined by
\begin{equation}\label{4}
N(\rho)=2\max\{0,-\lambda_{min}\},
\end{equation}
where $\lambda_{min}$  is the smallest eigenvalue of $\rho^{T_x}$,
$T_{x}$ stands for the partial transpose with respect to the subsystem $x$.

\section{Photon-induced entanglement in double cavities}

\begin{figure}[h]
\begin{center}
\end{center}
\caption{The schematic diagram of the model in this paper.}
\label{fig1}
\end{figure}

We consider a model consisting of two two-level atoms $A$ and $B$,
each interacting with a single-model near-resonant cavity field,
denoted $a$ and $b$ respectively, see Fig.\ref{fig1}. It is assumed that each
atom-cavity system is isolated and that the cavities are initially
in a entangled state while the atoms are,  different from the $X$
states \cite{yuting}, in a separable excited/ground state. The
dynamics of the model is characterized by the double J-C Hamiltonian
\begin{equation}\label{H1}
H=\frac{1}{2}{\omega}{\sigma}^{A}_{z}+\frac{1}{2}{\omega}{\sigma}^{B}_{z}
+g(a^{+}{\sigma}^{A}_{-}+a{\sigma}^{A}_{+})+g(b^{+}{\sigma}^{B}_{-}
+b{\sigma}^{B}_{+})+{\nu}{a}^{+}a+{\nu}{b}^{+}b,
\end{equation}
where $\nu$ is the field frequency, $\omega$ is the transition frequency between
the excited and the ground states of the atoms, $g$ is the coupling
constant between the cavity field and the atoms, $a$ and $a^{+}$
(resp. $b$ and $b^{+}$) are the field annihilation and creation operators associated
with the atom $A$'s (resp. $B$'s) cavity, and ${\sigma}_{\pm}$ are the spin-flip operators.
The eigenstates of the Hamiltonian (\ref{H1}) are products of the eigenstates
of the separate J-C system \cite{6}.
For simplicity, in the following we consider the case of zero detuning, $\omega=\nu$.
We denote $|\uparrow>$ (resp. $|\downarrow>$) the excited (resp. ground) state of the
atoms.

\subsubsection{Atoms initially in ground state}

We first study the case that cavities are initially
entangled while the two atoms are in the (separable) ground state,
$|\phi_{photon}>=\cos\alpha|01>+\sin\alpha|10>$,
$|\phi_{atom}>=|\downarrow\downarrow>$.
The initial state for the whole system is
$$
|\phi(0)>\,=\cos\alpha|\downarrow\downarrow01>+\sin\alpha|\downarrow\downarrow10>,
$$
where the physical Hilbert spaces from left to right correspond to
atoms $A$, $B$, photons $a$, $b$ respectively. In terms of the standard basis, the
state of the system at time $t$ can be written as
$$
|\phi(t)>=x_{1}(t)|\downarrow\downarrow01>+x_{2}(t)|\downarrow\downarrow10>
+x_{3}(t)|\downarrow\uparrow00>+x_{4}(t)|\uparrow\downarrow00>,
$$
where
$$
\ba{l}
 x_{1}(t)=\cos(gt)\cos\alpha,~~
 x_{2}(t)=\cos(gt)\sin\alpha,\\[2mm]
 x_{3}(t)=-i\sin(gt)\cos\alpha,~~
 x_{4}(t)=-i\sin(gt)\sin\alpha.
\ea
$$
The reduced density matrix $\rho^{AB}$ of two atoms can be obtained
by tracing out the photonic part of $|\phi(t)><\phi(t)|$. In the basis
$|\uparrow\uparrow>$, $|\uparrow\downarrow>$, $|\downarrow\uparrow>$, $|\downarrow\downarrow>$
it is of the form
\be\label{rhoAB1}
\rho^{AB}=\left(
\begin{array}{cccc}
0&0&0&0\\
0&|x_{4}|^{2}&x_{4}x^{\ast}_{3}&0\\
0&x_{3}x^{\ast}_{4}&|x_{3}|^{2}&0\\
0&0&0&|x_{1}|^{2}+|x_{2}|^{2}
\end{array}
\right).
\ee
The concurrence of the state (\ref{rhoAB1}) is given by
\be\label{CAB1}
C^{AB}=\sin^{2}(gt)|\sin2\alpha|.
\ee

\begin{figure}[h]
\begin{center}
\end{center}
\caption{Left figure: concurrence $C^{AB}$ with respect to the
initial state $|\phi(0)>$ as a function of time $t$ and parameter
$\alpha$; Right figure: the corresponding contour plot.}
\label{fig2}
\end{figure}

As shown in Fig.\ref{fig2}, one can see that the entanglement between the
atoms $A$ and $B$ varies periodically. The atoms keep disentangled
only when the photons are initially separable
($\alpha=0,~\pi/2,~\pi$). As long as the photons in two cavities are
entangled initially, the entanglement between atoms $A$ and $B$ can
be generated.

Similarly, by tracing out the two atoms part of $|\phi(t)><\phi(t)|$
we can obtain the reduced density matrix $\rho^{ab}$ with respect to the photons.
In the basis $|00>,~|01>,~|10>,~|11>$, we have
\be
\rho^{ab}=\left(
\begin{array}{cccc}
0&0&0&0\\
0&|x_{2}|^{2}&x_{2}x^{\ast}_{1}&0\\
0&x_{1}x^{\ast}_{2}&|x_{1}|^{2}&0\\
0&0&0&|x_{3}|^{2}+|x_{4}|^{2}
\end{array}
\right).
\ee
Therefore
\be\label{Cab1}
C^{ab}=\cos^{2}(gt)|\sin2\alpha|.
\ee

\begin{figure}[h]
\begin{center}
\end{center}
\caption{Left figure: concurrence $C^{ab}$ with respect to the
initial state $|\phi(0)>$ as a function of time $t$ and parameter
$\alpha$; Right figure: the corresponding contour plot.}
\label{fig3}
\end{figure}

From the Fig.\ref{fig3} we see that $C^{AB}$ increases when $C^{ab}$
decreases, and vice versa. The loss or gain of entanglement between
the two atoms is compensated by entanglement gain or loss between
the two photons. In fact, we have
\be\label{CABab}
C^{AB}+C^{ab}=|\sin2\alpha|.
\ee

The concurrence $C^{Aa}$and $C^{Ab}$ between the atoms and cavities can
be also obtained from their reduced density matrices,
$$
\ba{ll}
\rho^{Aa}=&|x_{3}|^{2}|\downarrow0><\downarrow0|+|x_{2}|^{2}|\downarrow1><\downarrow1|
+x_{2}x^{\ast}_{4}|\downarrow1><\uparrow0|\\
         &+x_{4}x^{\ast}_{2}|\uparrow0><\downarrow1|+|x_{4}|^{2}|\uparrow0><\uparrow0|
         +|x_{1}|^{2}|\downarrow0><\downarrow0|
\ea
$$
and
$$
\ba{ll}
 \rho^{Ab}=&|x_{3}|^{2}|\downarrow0><\downarrow0|+|x_{1}|^{2}|\downarrow1><\downarrow1|
 +x_{1}x^{\ast}_{4}|\downarrow1><\uparrow0|\\
 &+x_{4}x^{\ast}_{1}|\uparrow0><\downarrow1|
 +|x_{4}|^{2}|\uparrow0><\uparrow0|+|x_{2}|^{2}|\downarrow0><\downarrow0|,
\ea
$$
from which we have
\begin{eqnarray}\label{CAa1}
C^{Aa}=\sin^{2}\alpha|\sin(2gt)|,\\\label{CAb1}
C^{Ab}=|\sin2\alpha\sin(2gt)|/2.
\end{eqnarray}

From (\ref{CAB1}), (\ref{Cab1}), (\ref{CAa1}) and (\ref{CAb1}) we
see that there would be no ESD or ESB when the atoms are initially in the ground state.
An interesting phenomena here is the conservation of entanglement
between the atoms and photons (\ref{CABab}). There is a kind of entanglement transfer between
the atom pair and the photon pair. Namely the entanglement between the photons can be ``stored".
It gives a way to entangle
two remote atoms that are initially in a separable state.

\subsubsection{Atoms initially in excited state}

Now we consider the case that the atoms are initially in excited state.
The initial state for the system is of the form
$$
|\varphi(0)>\,=\cos\alpha|\uparrow\uparrow01>+\sin\alpha|\uparrow\uparrow10>.
$$
The state of the system at time $t$ is given by
$$
\ba{ll}
|\varphi(t)>\,=&y_{1}(t)|\uparrow\uparrow01>+y_{2}(t)|\uparrow\uparrow10>
+y_{3}(t)|\downarrow\uparrow11>+y_{4}(t)|\uparrow\downarrow02>\\&+y_{5}(t)|\downarrow\downarrow12>
+y_{6}(t)|\downarrow\uparrow20>+y_{6}(t)|\uparrow\downarrow11>+y_{8}(t)|\downarrow\downarrow21>,
\ea
$$
where according to Schr\"{o}dinger equation and the initial condition $|\varphi(0)>$,
$y_{1}(t)=\cos(gt)\cos({\sqrt{2}gt})\cos\alpha\,\, e^{-2it\nu}$,
$y_{2}(t)=\cos(gt)\cos({\sqrt{2}gt})\sin\alpha\,\, e^{-2it\nu}$,
$y_{3}(t)=-i\sin(gt)\cos({\sqrt{2}gt})\cos\alpha\,\, e^{-2it\nu}$,
$y_{4}(t)=-i\cos(gt)\sin({\sqrt{2}gt})\cos\alpha\,\, e^{-2it\nu}$,
$y_{5}(t)=-\sin(gt)\sin({\sqrt{2}gt})\cos\alpha\,\, e^{-2it\nu}$,
$y_{6}(t)=-i\cos(gt)\sin({\sqrt{2}gt})\sin\alpha\,\, e^{-2it\nu}$,
$y_{7}(t)=-i\sin(gt)\cos({\sqrt{2}gt})\sin\alpha\,\, e^{-2it\nu}$,
$y_{8}(t)=-\sin(gt)\sin({\sqrt{2}gt})\sin\alpha\,\, e^{-2it\nu}$.

The reduced density matrix for atoms is given by
\be
\rho^{AB}=\left(
\begin{array}{cccc}
|x_{1}|^{2}+|x_{2}|^{2}&0&0&0\\
0&|x_{4}|^{2}+|x_{7}|^{2}&x_{7}x^{\ast}_{3}&0\\
0&x_{3}x^{\ast}_{7}&|x_{3}|^{2}+|x_{6}|^{2}&0\\
0&0&0&|x_{5}|^{2}+|x_{8}|^{2}
\end{array}
\right)
\ee
and the corresponding concurrence is given by $C^{AB}=2\max\{0,f(t)\}$ , where
$$
f(t)=\frac{1}{2}\sin^{2}(gt)\cos^{2}(\sqrt{2}gt)|\sin2\alpha|-\frac{1}{4}|\sin(2gt)\sin(2\sqrt{2}gt)|.
$$

\begin{figure}[h]
\begin{center}
\end{center}
\caption{Left figure: concurrence
$C^{AB}$ with respect to the initial state $|\varphi(0)>$ as a function of
time $t$ and parameter $\alpha$; Right figure: the corresponding contour plot.}
\label{fig5}
\end{figure}

From Fig.\ref{fig5} we can see the novel entanglement sudden death and
sudden birth phenomena \cite{yuting,12}. Moreover the
length of the time interval for the zero entanglement is not
dependent on the degree of entanglement of the initial state, in
consistent with the result of the double J-C model \cite{13}.

The atom-cavity photon system is now a qubit-qutrit one.
The corresponding reduced density
matrices $\rho^{Aa}$ and $\rho^{Ab}$ are $6\times6$ ones:
\be\rho^{Aa}=\left(
\begin{array}{cccccc}
0&0&0&0&0&0\\
0&|x_{2}|^{2}+|x_{7}|^{2}&0&x_{2}x^{\ast}_{6}+x_{7}x^{\ast}_{8}&0&0\\
0&0&|x_{1}|^{2}+|x_{4}|^{2}&0&x_{1}x^{\ast}_{3}+x_{4}x^{\ast}_{5}&0\\
0&x_{6}x^{\ast}_{2}+x_{8}x^{\ast}_{7}&0&|x_{6}|^{2}+|x_{8}|^{2}&0&0\\
0&0&x_{3}x^{\ast}_{1}+x_{5}x^{\ast}_{4}&0&|x_{3}|^{2}+|x_{5}|^{2}&0\\
0&0&0&0&0&0\\
\end{array}
\right),
\ee
\be\label{pp}
\rho^{Ab}=\left(
\begin{array}{cccccc}
|x_{4}|^{2}&0&0&0&0&0\\
0&|x_{1}|^{2}+|x_{7}|^{2}&0&x_{7}x^{\ast}_{5}&0&0\\
0&0&|x_{2}|^{2}&0&x_{2}x^{\ast}_{3}&0\\
0&x_{5}x^{\ast}_{7}&0&|x_{5}|^{2}&0&0\\
0&0&x_{3}x^{\ast}_{2}&0&|x_{3}|^{2}+|x_{8}|^{2}&0\\
0&0&0&0&0&|x_{6}|^{2}
\end{array}
\right).
\ee
We use negativity to quantify the entanglement between atom $A$ and cavity $a$ ($b$).
According to equation (\ref{4}), we have
$$
\ba{ll}
N^{Aa}=&\sqrt{4\sin^{2}(gt)\cos^{2}(gt)\cos^{4}\alpha+\cos^{4}(\sqrt{2}gt)\sin^{4}\alpha}
-\cos^{2}(\sqrt{2}gt)\sin^{2}\alpha\\
&+\sqrt{4\sin^{2}(\sqrt{2}gt)\cos^{2}(\sqrt{2}gt)\sin^{4}\alpha+\sin^{4}(gt)\cos^{4}\alpha}
-\cos^{2}\alpha\sin^{2}(gt)
\ea
$$
and
$$
N^{Ab}=2\max\{0,-\lambda_{min}\},
$$
where
$$
\ba{ll}
\lambda_{min}=&\frac{1}{2}
\left\{\cos^{2}(\sqrt{2}gt)\cos^{2}\alpha\sin^{2}(gt)
+\cos^{2}(gt)\cos^{2}\alpha\sin^{2}(\sqrt{2}gt)\right.\\
&+\sin^{2}(gt)\sin^{2}(\sqrt{2}gt)\sin^{2}\alpha-
[\cos^{4}(\sqrt{2}gt)\cos^{4}\alpha\sin^{4}(gt)\\
&+\cos^{4}(gt)\cos^{4}\alpha\sin^{4}(\sqrt{2}gt)
+\sin^{4}(gt)\sin^{4}(\sqrt{2}gt)\sin^{4}\alpha\\
&+6\sin^{4}(gt)\sin^{2}(\sqrt{2}gt)\cos^{2}(\sqrt{2}gt)\sin^{2}\alpha\cos^{2}\alpha\\
&-2\cos^{4}\alpha\sin^{2}(gt)\cos^{2}(gt)\sin^{2}(\sqrt{2}gt)\cos^{2}(\sqrt{2}gt)\\
&\left.-2\sin^{4}(\sqrt{2}gt)\sin^{2}(gt)\cos^{2}(gt)\sin^{2}\alpha\cos^{2}\alpha]^{\frac{1}{2}}\right\}.
\ea
$$

From Fig.\ref{fig6} we see that the entanglement between atom $A$ and
photon $a$ varies continuously with time. Nevertheless the
entanglement between atom $A$ and photon $b$, see Fig.\ref{fig7}, has
again entanglement sudden death and sudden birth phenomena.

\begin{figure}[h]
\begin{center}
\end{center}
\caption{Left figure: concurrence
$N^{Aa}$ with respect to the initial state $|\varphi(0)>$ as a function of
time $t$ and parameter $\alpha$; Right figure: the corresponding contour plot.}
\label{fig6}
\end{figure}

\begin{figure}[h]
\begin{center}
\end{center}
\caption{Left figure: concurrence
$N^{Ab}$ with respect to the initial state $|\varphi(0)>$ as a function of
time $t$ and parameter $\alpha$; Right figure: the corresponding contour plot.}
\label{fig7}
\end{figure}

In Fig.\ref{fig7} we also see the ESD and ESB effects between the atom $A$
and photon $b$. While the entanglement between
atom $A$ and the adjacent cavity $a$ has no such effects, similar to the case
in last subsection when the two atoms are initially in a separable ground state.

\subsubsection{Entanglement enhancement by quantum Zeno effect}

From (\ref{CAb1}) we see that the entanglement between atom $A$ and remote
photon $b$ reaches only to the maximum 0.5 for suitable initial condition $\alpha$.
To protect and enhance entanglement quantum Zeno effect has been taken into account \cite{18,19}.
Below we show that if the dynamics is
controlled by quantum Zeno effect, the maximally entanglement $1$ can be attained.

Set $P_{B}=I_{A}\otimes|\downarrow><\downarrow|\otimes I_{a}\otimes I_{b}$,
which acts on the subsystem $B$, projecting the state to its initial one.
Under the evolution with $N$ projective measurements on $B$, one obtains
$$
\ba{rcl}
|\phi(t)>_{N}&=&(P_{B}e^{-iHt/N\hbar})^N|\phi(0)>\\
&=&\cos\alpha\cos^{N}(\frac{gt}{N})|\downarrow\downarrow01>+
\sin\alpha\cos(gt)|\downarrow\downarrow10>-i\sin\alpha\sin(gt)|\uparrow\downarrow00>.
\ea
$$
Under the limit $N\rightarrow\infty$, we get
$$
\lim_{N\rightarrow\infty}C^{Ab}_{N}=|\sin(2\alpha)\sin(gt)|.
$$

Obviously now the entanglement between the atom
$A$ and the remote photon $b$ is enhanced and reaches the maximum $1$
for suitable initial states, see Fig.\ref{fig4} for $\alpha=\frac{\pi}{4}$.

\begin{figure}[h]
\begin{center}
\end{center}
\caption{Solid line: the entanglement $C^{Ab}$ under free
dynamics. Dashing line: the entanglement $C^{Ab}$ under
projective measurements.}
\label{fig4}
\end{figure}

From Fig.\ref{fig7} we also see that when the atoms were initially
in exited state, the maximal entanglement between the atom $A$
and the remote photon $b$ is only $0.2$. By using quantum Zeno effect,
after $N$ projective measurements on $B$, we can get (for simplicity
we take $w=g$)
$$
\ba{ll}
 |\varphi(t)>_{N}=&e^{\frac{-3igt}{2}}\cos\alpha\cos(gt)
\left[\cos(\frac{3gt}{2N})-\frac{i}{3}\sin(\frac{3gt}{2N})]^{N}|\uparrow\uparrow01>\right.\\
&-ie^{\frac{-3igt}{2}}\cos\alpha\sin(gt)
[\cos(\frac{3gt}{2N})-\frac{i}{3}\sin(\frac{3gt}{2N})]^{N}|\downarrow\uparrow11>\\
&+e^{\frac{-3igt}{2}}\sin\alpha\cos^{N}(\frac{gt}{N})
[-\frac{1}{3}\sinh(\frac{3igt}{2})+\cosh(\frac{3igt}{2})]|\uparrow\uparrow10>\\
&\left.-\frac{2\sqrt[]{2}}{3}e^{\frac{-3igt}{2}}\sin\alpha\cos^{N}(\frac{gt}{N})
\sinh(\frac{3igt}{2})|\downarrow\uparrow20>\right].
 \ea
$$
and
$$
\lim_{N\rightarrow\infty}N^{Ab}_{N}=2\max\{0,-f(t)_{min}\},
$$
where
$$
\ba{ll} f(t)_{min}=&\frac{1}{2}
\left\{\cos^{2}\alpha\cos^{2}(gt)+\frac{8}{9}\sin^{2}\alpha\sin^{2}(\frac{3gt}{2})\right.\\
&-[\cos^{4}\alpha\cos^{4}(gt)+\frac{64}{81}\sin^{4}\alpha\sin^{4}(\frac{3gt}{2})\\
&+4\sin^{2}\alpha\cos^{2}\alpha\sin^{2}(gt)[\cos^{2}(\frac{3gt}{2})+\frac{1}{9}\sin^{2}(\frac{3gt}{2})]\\
&-\left.\frac{16}{9}\sin^{2}\alpha\cos^{2}\alpha\cos^{2}(gt)\sin^{2}(\frac{3gt}{2})]^{\frac{1}{2}}\}\right\}.
\ea
$$

From the Fig.\ref{fig8} we can see that the entanglement between
atom $A$ and the remote photon $b$ has been improved.

\begin{figure}[h]
\begin{center}
\end{center}
\caption{Solid line: $N^{Ab}$ under free
dynamics; Dashed line: $N^{Ab}$ under
projective measurements for $\alpha=\frac{\pi}{4}$.}
\label{fig8}
\end{figure}

\section{Discussions}

We have investigated entanglement
evolution among atoms and photons in cavities in terms of
Jaynes-Cummings model. It has been shown that for a pair of
atoms in separated cavities and in a separable state, if a pair of entangled photons
are introduced into the cavities, the entanglement between the two atoms can be established.
This gives away to entangle two remote atoms on the one hand. It also gives rise
to a kind of entanglement storage on the another hand.

In particular, when two atoms are initially in the exited state,
there exit both entanglement sudden death and entanglement
sudden birth phenomena between the atoms, and between the atom in one cavity
and the photon in another cavity.
Interestingly the time interval of zero entanglement is independent of the
entanglement of the initial state of the photons.

Moreover the maximal entanglement attained between the atoms depends
on the maximal entanglement of the photons. If the photons are
initially maximally entangled, then the atoms can evolve into
maximally entangled states. Nevertheless the entanglement between
the atom in one cavity and the photon in another cavity is
relatively small. We have shown that their entanglement can be
enhanced in terms of quantum Zeno effect. For the case that two
atoms were initially in ground state, this kind of entanglement can
be even improved to be the maximal one.

\bigskip
\noindent{\bf Acknowledgments}\, This work was supported by the NSFC
10875081, KZ200810028013 and PHR201007107.

\end{document}